\documentclass{iauc}
\usepackage{graphicx}

\title[SDI Survey] 
{A Survey of Close, Young Stars with the Simultaneous Differential 
Imager (SDI) at the VLT and MMT}

\author[Biller et al.]   
{Beth A. Biller$^1$, Laird M. Close$^1$, Elena Masciadri$^{2}$, 
Rainer Lenzen$^{2}$, Wolfgang Brandner$^{2}$, Donald McCarthy$^1$, 
Thomas Henning$^2$, Eric Nielsen$^1$, and Markus Hartung$^3$ }

\affiliation{$^1$Steward Observatory, University of Arizona, Tucson, AZ 85721
\break email: bbiller@as.arizona.edu\\[\affilskip]
$^2$ Max-Planck-Institut f\"ur Astronomie, K\"onigstuhl 17, 69117 
Heidelberg, Germany \break
$^3$ European Southern Observatory, Alonso de Cordova 3107, 
Santiago 19, Chile \break}

\pubyear{2005}
\volume{200} 
\pagerange{1--7}
\date{?? and in revised form ??}
\jname{Direct Imaging of Exoplanets: Science and Techniques}
\editors{C. Aime and F. Vakili, Eds.}
\begin{document}

\maketitle

\begin{abstract}
We discuss the preliminary results of a survey of young ($<$300 Myr),
close ($<$50 pc) stars with the Simultaneous Differential Extrasolar
Planet Imager (SDI)
implemented at the VLT and the MMT.  SDI uses a double Wollaston prism and 
a quad filter to take 4 identical images simultaneously
at 3 wavelengths surrounding the 1.62 $\mu$m methane bandhead found in 
the spectrum of cool brown dwarfs and gas giants.  By performing 
a difference of images in these filters, speckle noise from
the primary can be significantly attenuated, resulting in photon noise limited 
data.  In our survey data, we achieved H band contrasts
$>$25000 (5$\sigma$ $\Delta$F1(1.575$\mu$m)$>$10 mag, 
$\Delta$H$>$11.5 mag for a T6 spectral
type) at a separation of 0.5" from
the primary star.  With this degree of attenuation, we should be able to 
image (5$\sigma$ detection)
a 2-4 Jupiter mass planet at 5 AU around a 30 Myr star at 10 pc.
We are currently completing our survey of young, nearby stars.  We have 
obtained complete
datasets for 35 stars in the southern sky (VLT) 
and 7 stars in the northern sky (MMT).
We believe that our SDI images are the highest contrast astronomical 
images ever made from ground or space for methane rich companions.
\end{abstract}

\firstsection 
\section{Introduction}

Direct detection of extrasolar giant planets is extremely difficult.  
Giant gas planets seen in reflected light are $>$20 magnitudes fainter 
than their primary stars and likely 
lie within $\sim$1'' of their primary stars.  
The problem is slightly easier with younger, hotter planets -- 100 Myr old 
extra-solar planets are 10$^{4-7}$ times more self-luminous than old (5 Gyr) 
extra-solar planets, whereas their primary stars are only slightly 
(2-5 times) brighter at early ages.  In theory, adaptive optics (AO) 
systems that are ``photon noise limited'' can detect an object up to 
10$^5$ times fainter than 
its primary at separations $>$1''.  However, numerous surveys for 
extrasolar planets using large telescopes with AO systems have yielded useful
limits but few confirmed giant planet candidates 
(Kaisler et al. 2003, Masciadri et al. 2005, Chauvin et al. 2005, Neuh\"auser
et al. 2005).  

AO surveys for young extrasolar planets 
only address half of the difficulty of direct detection -- the contrast limit
problem.  Beyond the contrast limit problem, all AO systems suffer from a 
limiting ``speckle noise'' floor (Racine et al. 1999).  Within 1'' 
of the primary star, the field is filled with speckles left over from 
instrumental features and residual atmospheric turbulence after adaptive 
optics correction.  These speckles vary as a function of time and color.  
For photon noise limited data, the signal to noise S/N increases as
 t$^{0.5}$, 
where t is the exposure time.  For speckle-noise limited data, the S/N does 
not increase with time past a specific speckle-noise floor (limiting 
contrasts to $\sim$10$^3$ at 0.5'').  This speckle-noise floor is 
considerably above the photon noise limit and makes planet detection very 
difficult.  Interestingly, space telescopes such as HST also suffer from a 
similar limiting speckle-noise floor due to imperfect optics and 
``breathing'' (Schneider et al. 2003).  Direct detection of 
extrasolar giant planets requires special new instrumentation to suppress 
this speckle noise floor and produce photon noise limited images.  The VLT, 
Keck, Subaru, and Gemini are all currently developing dedicated planet-finding
cameras which exploit these new instrumentational approaches for speckle
suppression.  The Simultaneous Differential Imager (SDI), 
which our team built and installed at the 
VLT and MMT (see Biller et al., this conference), 
is one of the first dedicated planet-finding optical devices to come online
on a large telescope.
 
Simultaneous Differential Imaging is an instrumental method which can be 
used to calibrate and remove the ``speckle noise'' in AO images, while 
also isolating the planetary light from the starlight.  This method was 
pioneered by Racine et al. (1999), Marois et al. (2000), Marois et al. 
(2002), and Marois et al. (2005).  
It exploits the fact that all cool (T$_{eff}$ $<$1200 K) 
extra-solar giant planets have strong CH$_4$ (methane) absorption 
redwards of 1.62 $\mu$m in the H band infrared atmospheric window 
(Burrows et al. 2001, Burrows et al. 2003).  Our SDI device obtains four 
images of a star simultaneously through three slightly different narrowband
 filters (sampling both inside and outside of the CH$_4$ features).  These 
images are then differenced.  This subtracts out the halo and speckles from 
the bright star to reveal any massive extrasolar planets orbiting that star.  
Since a massive planetary companion will be brightest in one filter and 
absorbed in the rest, while the star is bright in all three, 
a difference can be chosen which subtracts out the star's light 
and reveals the light from the companion.  Thus, SDI also helps 
eliminate the large contrast difference between the star and substellar 
companions (Close et al. 2005; Lenzen et al. 2004; Lenzen et al. 2005)  
The SDI device has 
already produced a number of important scientific results: the discovery 
of AB Dor C (Close et al. 2005) which is the tightest (0.16'') 
low mass companion detected by direct imaging, detailed 
surface maps of Titan 
(Hartung et al. 2004), the discovery of $\epsilon$ Indi Ba-Bb, the 
nearest binary brown dwarf (McCaughrean et al. 2005), and evidence 
of orbital motion for Gl 86B, the first known 
white dwarf companion to an exoplanet host star
(Mugrauer and Neuh\"auser 2005).  

\section{The SDI survey}
 \begin{figure}
   \begin{center}
   \begin{tabular}{c}	
	\includegraphics[height=3.5cm]{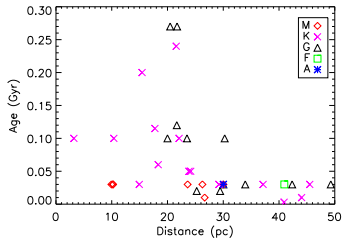}	
	\end{tabular}
	\end{center}
	\caption[Age vs. Distance]
	{\label{fig:agevsdist}Age vs. Distance for our observed sample stars}
	\end{figure}	

We are currently completing a survey with the SDI device of $\sim$50 
young ($<$300 Myr), nearby ($<$50 pc) stars.  
Stars were chosen based on strong lithium absorption 
features (our best targets have Li equivalent widths of $>$100 m{\AA} from the 
Li 6707 {\AA} line, corresponding to age $<$ 100 Myr) and accurate 
Hipparcos parallax measurements (parallaxes of $>$ 0.02'', corresponding to 
distances $<$ 50 pc).  Complete datasets have been acquired for 42 stars 
total -- 35 stars in the southern sky (VLT) 
and 7 stars in the northern sky (MMT).    
Ages, distances, and spectral types of observed objects are presented in 
Fig.~\ref{fig:agevsdist}.  The ``average'' survey object is a late K star with
an age of 120 Myr and at a distance of 26 pc.

Based on realistic scaling laws for the semimajor 
axis, eccentricity, mass power law, and luminosity of extrasolar planets 
(statistics from Marcy et al. 2003 and Lineweaver and Grether 2003, 
masses and ages from Burrows et al. 2003) and scaling to semimajor axes 
$>$ 5 AU, we can determine a rough detection probability for each program 
star (read Nielsen et al. this conference for more detail).  
Our program stars have average 
detection probabilities of $\sim$10-20$\%$.  Integrating over the probability
distribution of our program stars yields $\sim$4 likely detections in 50 stars.

A number of tentative candidate extrasolar planets 
(objects which showed CH$_4$ 
absorption at appropriate separations from the primary) have been
identified so far -- followup observations of these candidates are scheduled 
to confirm if these candidates are real and 
share common proper motion with their 
parent stars.  Low resolution 
follow-up CONICA grism spectroscopy will be performed on any confirmed 
candidate once the exact planet location is confirmed.  With our excellent 
spatial resolution, we can determine an extremely accurate offset between 
the planet and thus will be able to precisely align a narrow 0.1'' slit on 
the planet.  See Close et al. 2005 and Nielsen et al. 2005 for details on 
spectral reduction.

\section{An Example Dataset}
   \begin{figure}
   \begin{center}
   \begin{tabular}{ccc}
   \includegraphics[height=3.2cm]{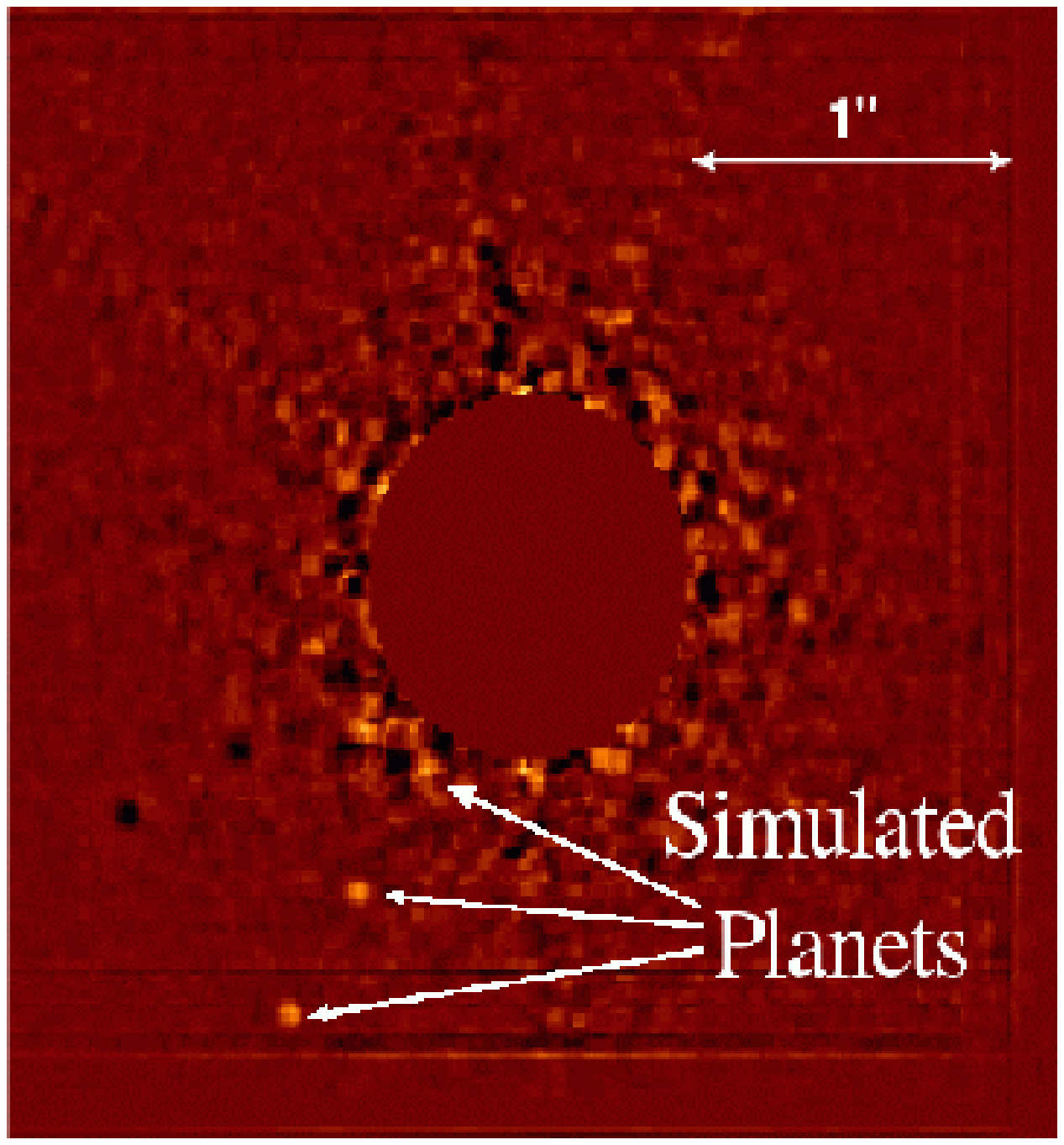} &
   \includegraphics[height=3.2cm]{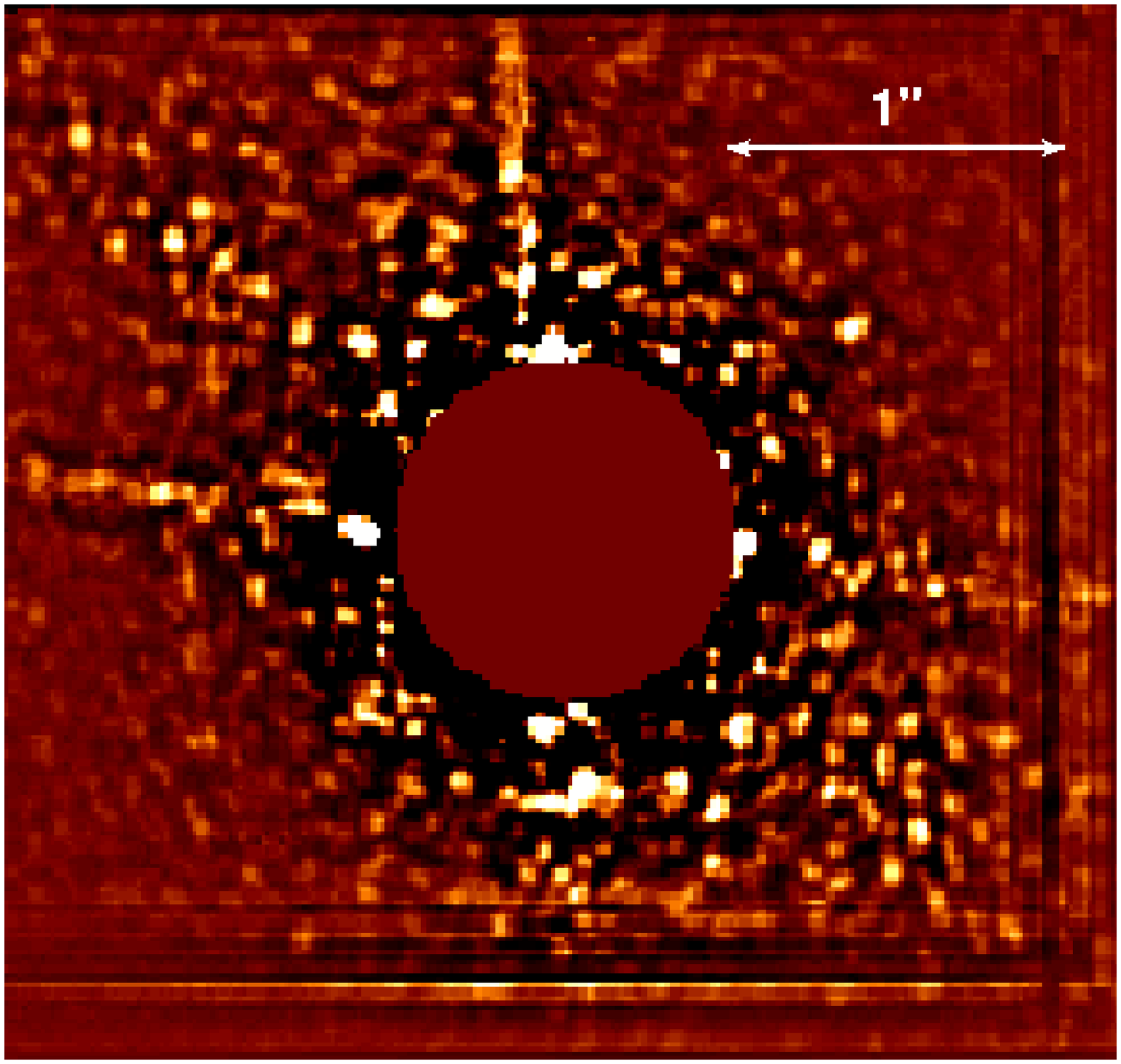} &
   \includegraphics[height=3.2cm]{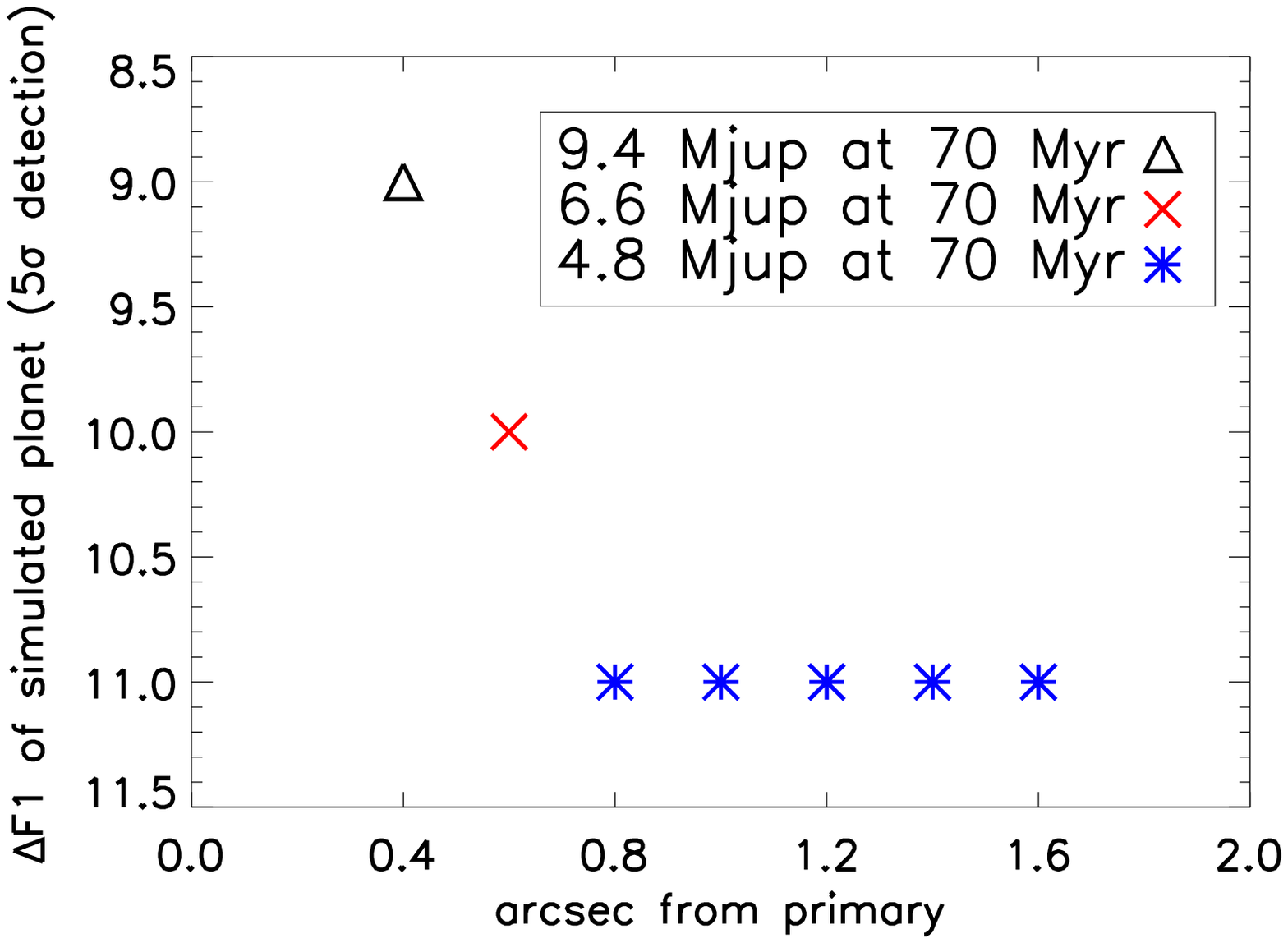}\\
   \end{tabular}
   \end{center}
   \caption[Reduced VLT SDI data]
   { \label{fig:SDIRED} {\bf Left:} A complete reduced dataset 
(40 minutes of data at a series of rotator angles -- 
0$^{\circ}$, 33$^{\circ}$, 33$^{\circ}$, 0$^{\circ}$) from the VLT SDI device.
Simulated planets have been added at separations of 
0.55, 0.85, and 1.35'' from the primary, with $\Delta$F1(1.575$\mu$m) = 10 mag 
(attenuation in magnitudes in the 1.575 $\mu$m 
F1 filter) fainter than the primary.
  These planets are scaled from unsaturated images of the example star 
taken right before the example dataset (and have fluxes and 
photon noise in each filter 
appropriate for a T6 object).  Past 0.7'', the simulated planets are 
detected with S/N $>$ 10.
{\bf Center:} Standard AO data reduction of the same
dataset.  
Filter images have been coadded (rather than subtracted), 
flat-fielded, sky-subtracted, and 
unsharp-masked.  Simulated planets have been added with the 
same properties and at the same separations as before.  None of the simulated
planets are detected in the standard AO reduction.  Additionally, numerous
bright super speckles remain in the field.
{\bf Right:} Minimum Detectable Planet Mass (5$\sigma$
detection vs. Separation for this example dataset (Case B).  To 
determine minimum detectable planet mass as a function of separation, 
we inserted and then attempted to retrieve simulated planets with a variety 
of separations and $\Delta$F1 (5$\sigma$ 
noise level in the 1.575 $\mu$m
F1 filter) contrasts.  $\Delta$F1 contrasts were 
translated into planet masses using the models of Burrows et al. 2003.  For 
this particular star, we can detect a 5 M$_{Jup}$ planet 12 AU from the star.
}
\end{figure}

 \begin{figure}
   \begin{center}
   \begin{tabular}{cc}
   \includegraphics[height=5cm]{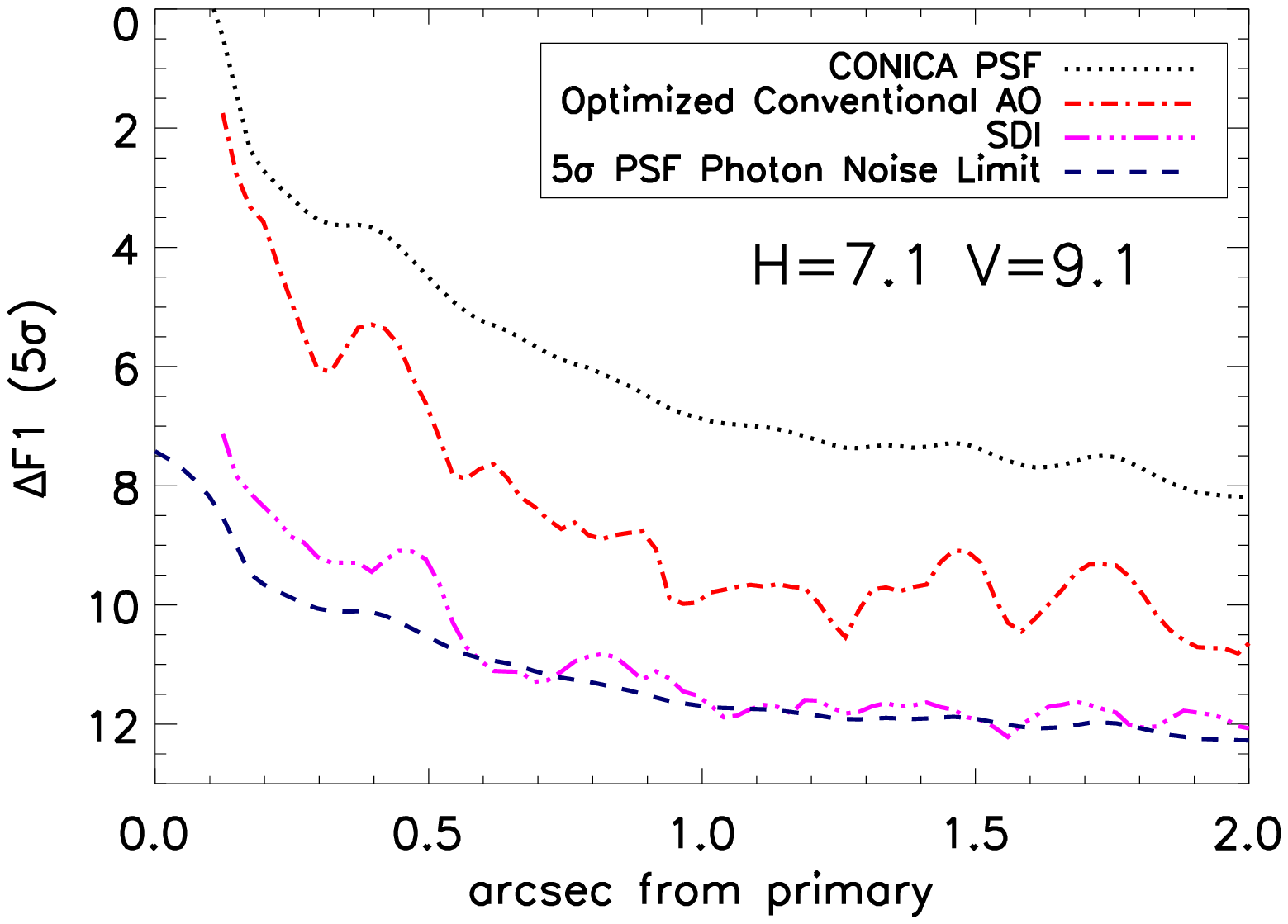} &
   \includegraphics[height=5cm]{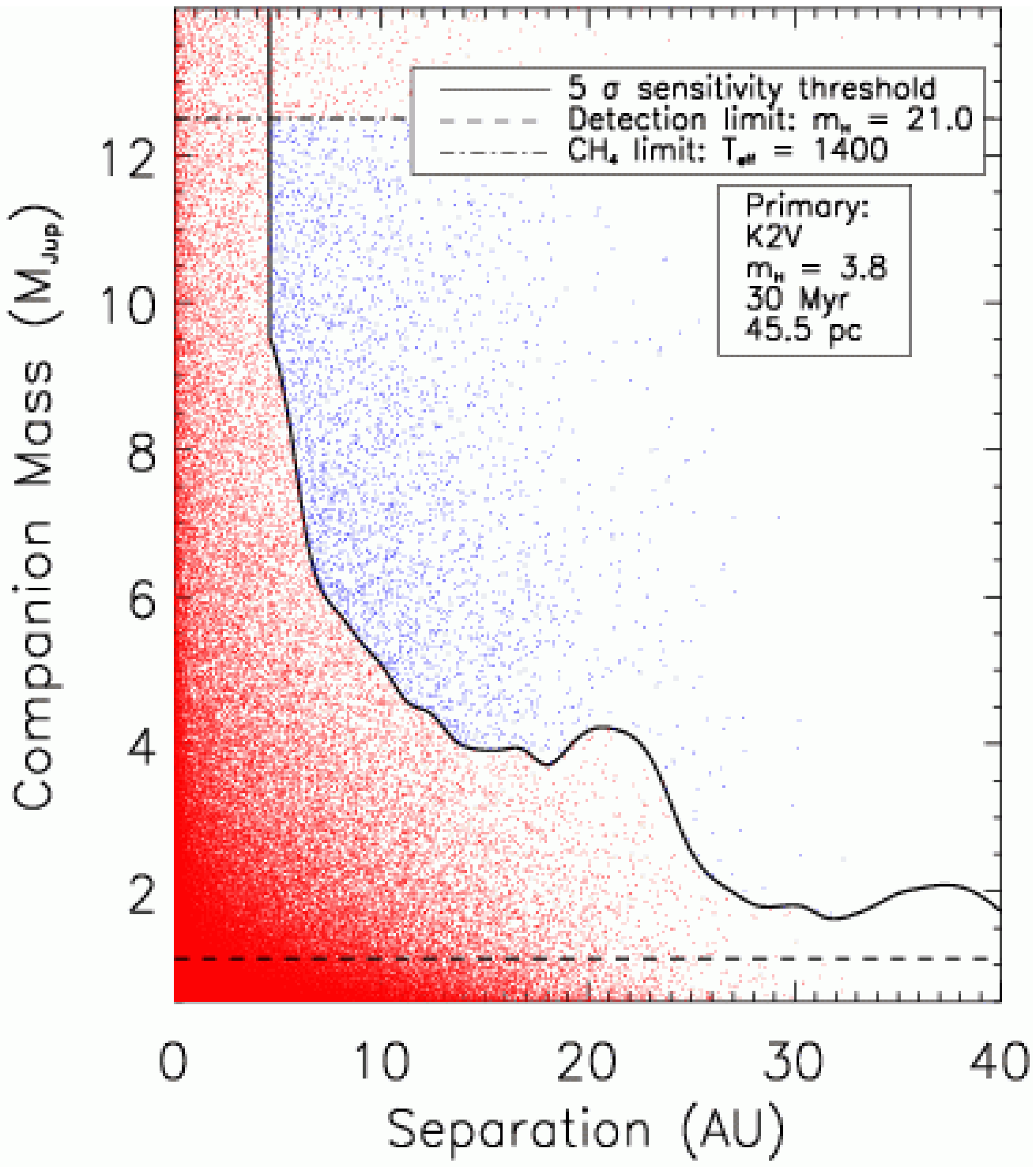} \\ 
   \includegraphics[height=5cm]{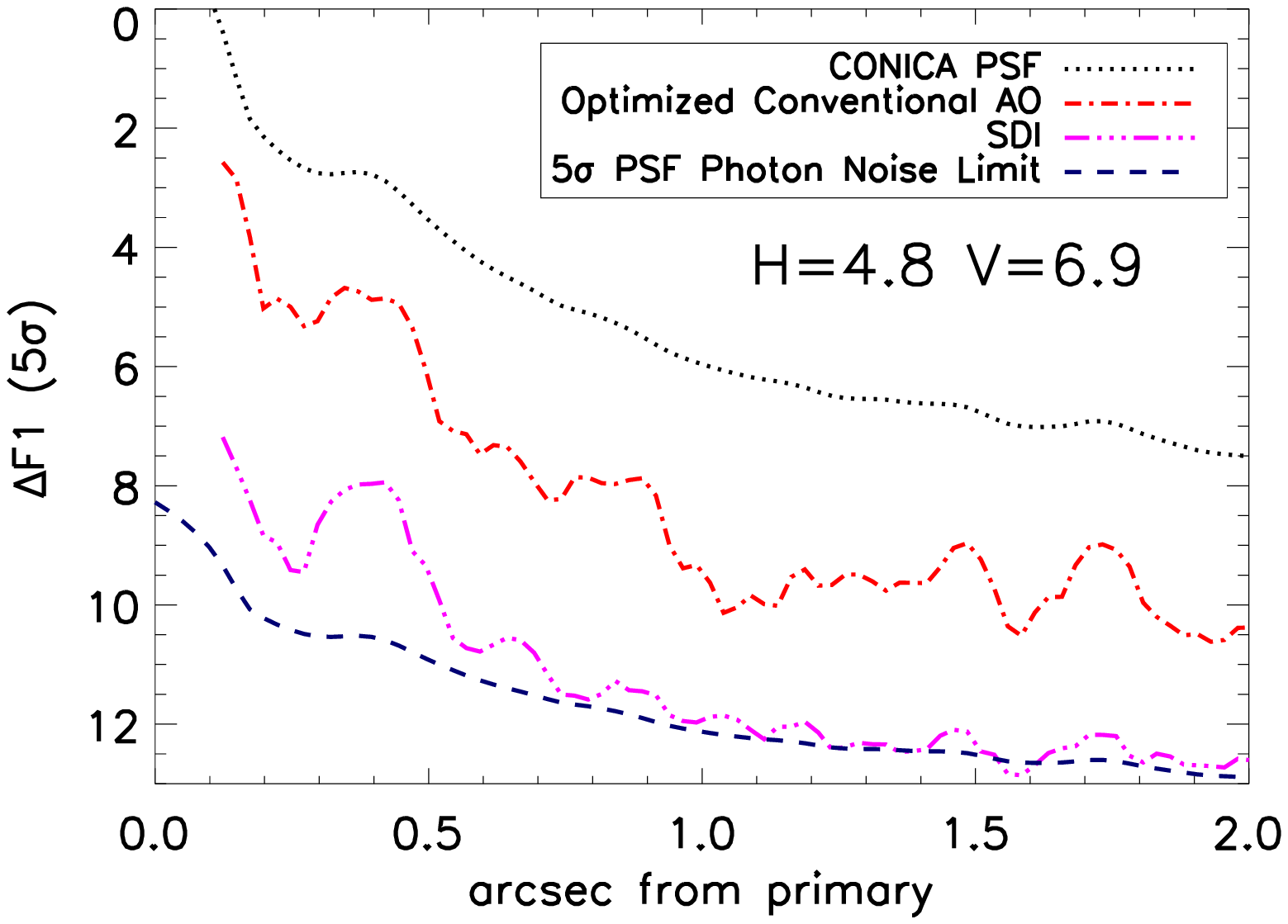} &
   \includegraphics[height=5cm]{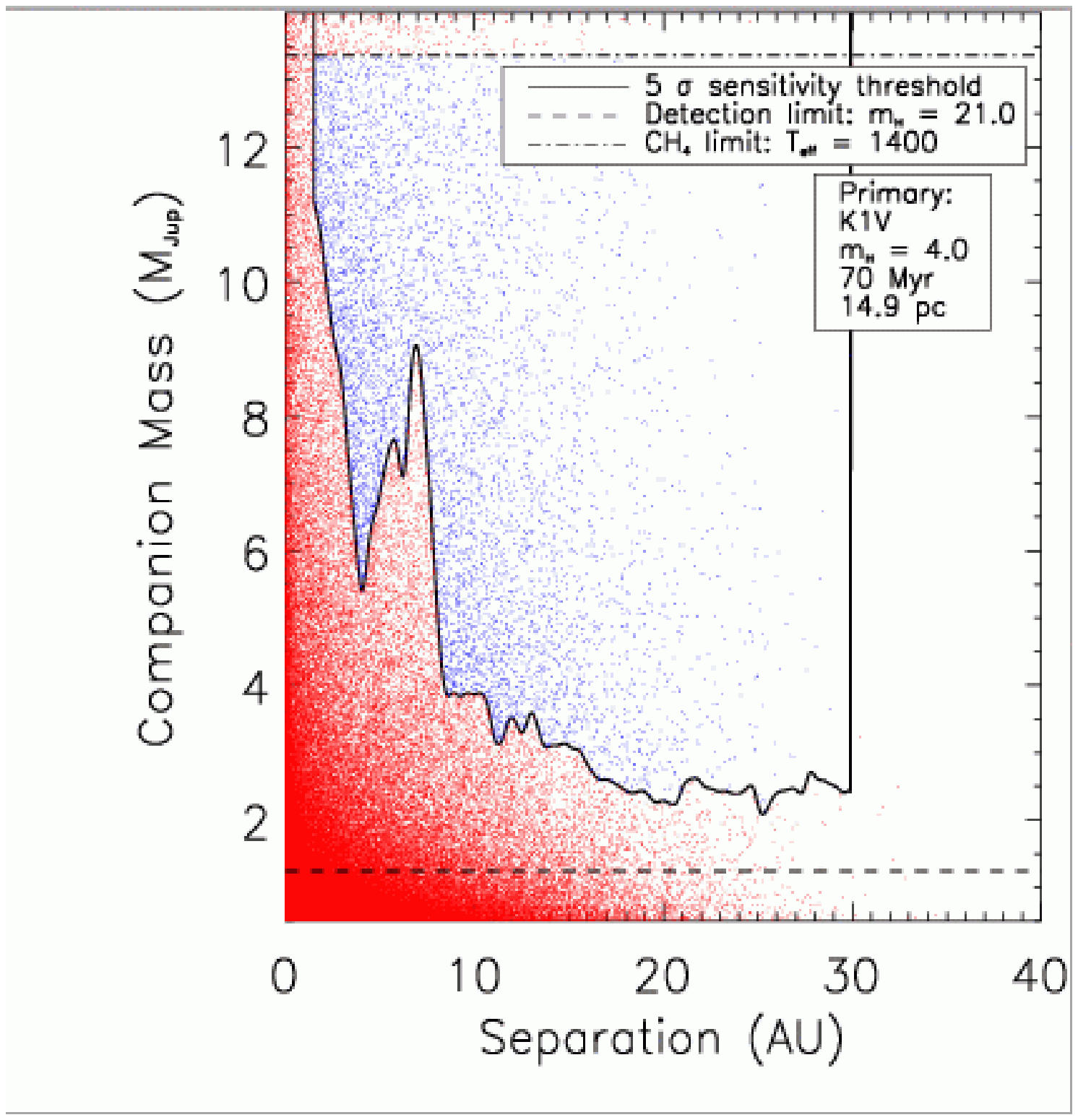} \\ 
   \includegraphics[height=5cm]{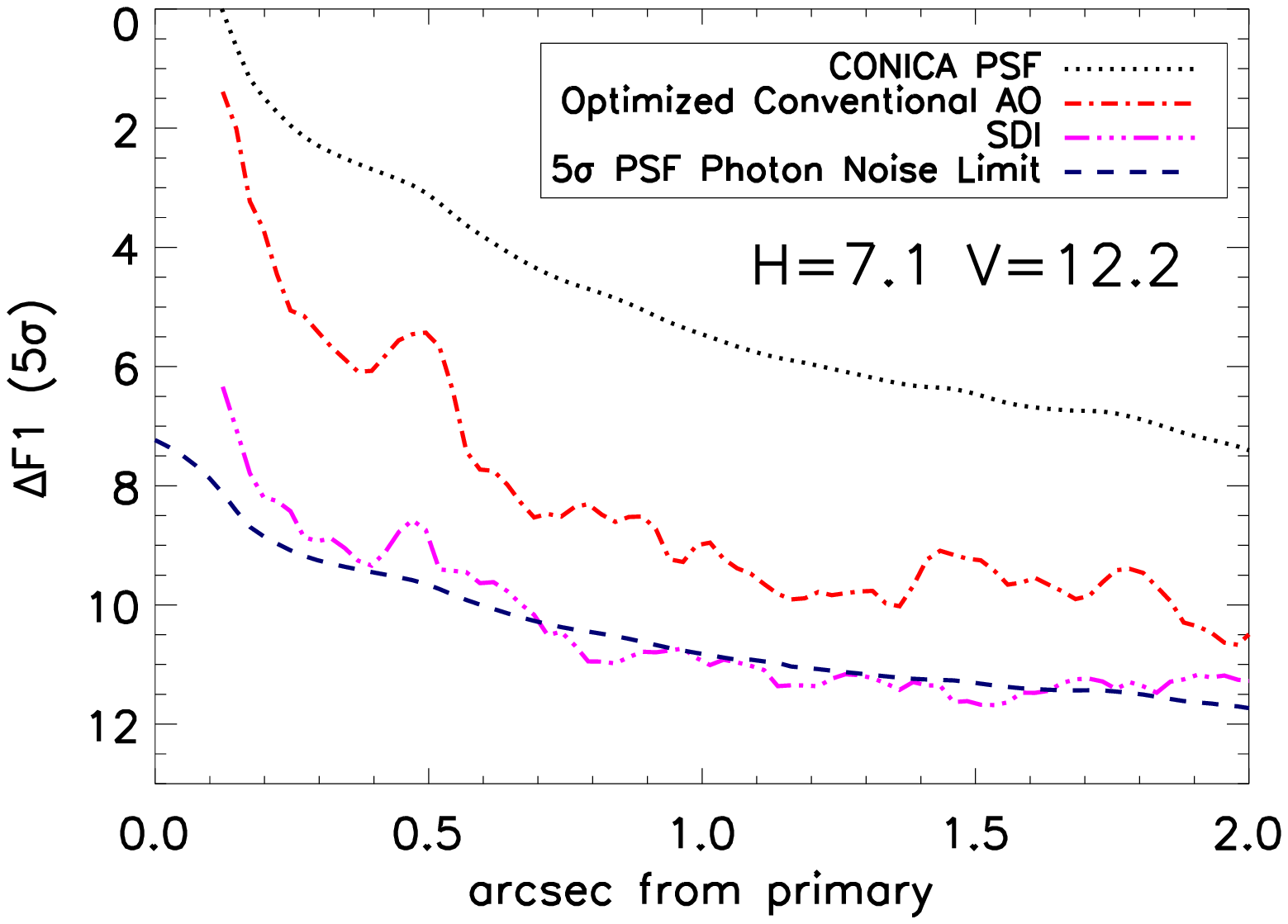} & 
   \includegraphics[height=5cm]{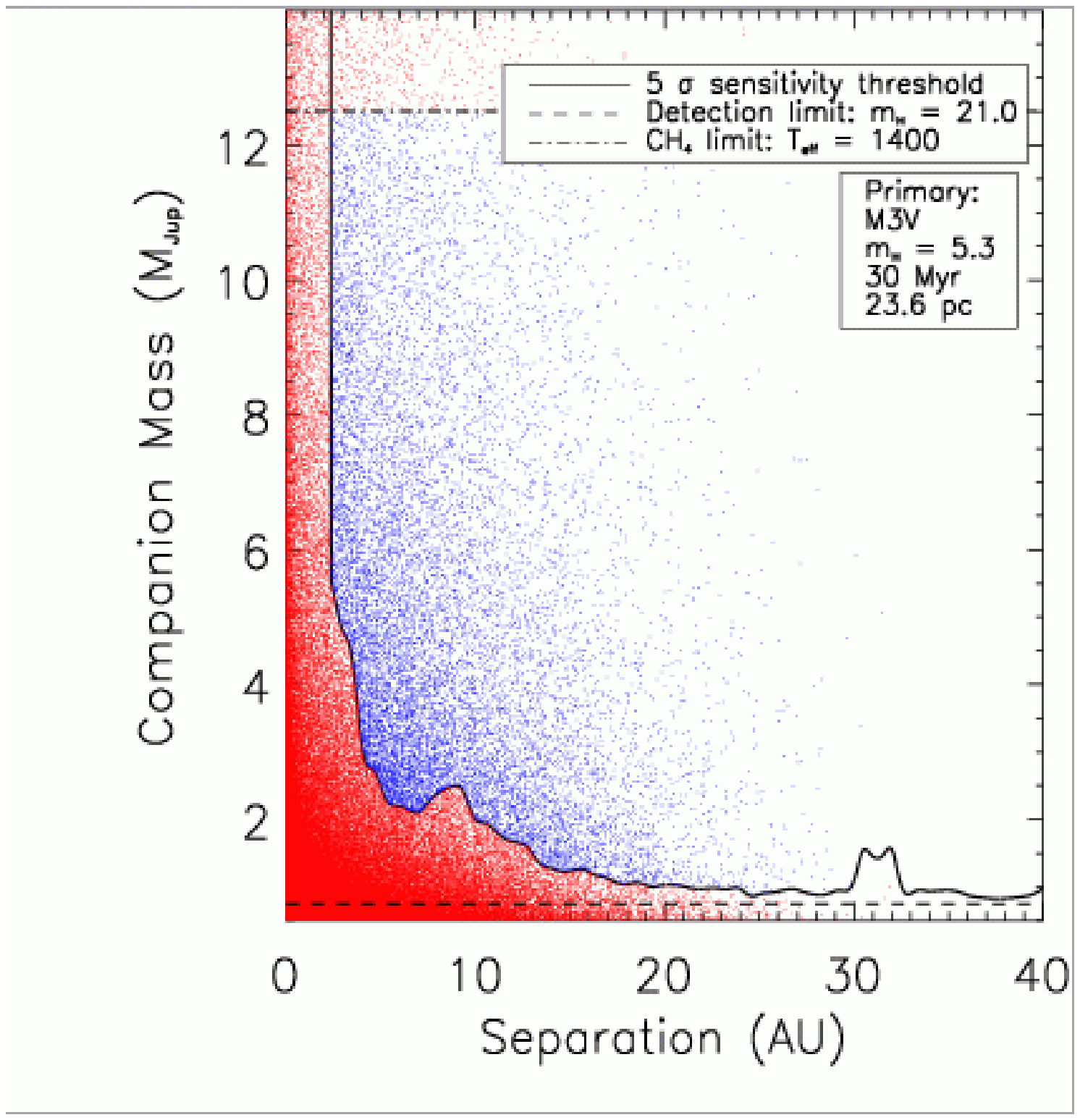} \\ 
   \end{tabular}
   \end{center}
   \caption[$\Delta$F1 vs. Separation]
2
   { \label{fig:deltaH} {\bf Top:} Case A, {\bf Middle:} Case B, {\bf Bottom:} Case C.
{\bf Left:} $\Delta$F1 (5$\sigma$ attenuation in magnitudes in the 1.575 $\mu$m
F1 filter) vs. Separation for 40 minutes of VLT SDI 
data for each example star. The top curve is the AO PSF.  The next 
curve is the ``classical AO PSF'' unsharp masked.  The third curve down is 
40 minutes of SDI data taken at two different position angles and subtracted
(0$^{\circ}$ data -33$^{\circ}$ data).  
The last curve is the theoretical contrast
limit due to photon-noise.  For each case, 
at star-companion separations $>$0.5'', 
we are photon-noise limited and achieve H band star to planet 
contrasts $>$25000 (5$\sigma$ $\Delta$F1(1.575$\mu$m)$>$10 mag, 
$\Delta$H$>$11.5 mag for a T6 spectral type)
{\bf Right:}  Minimum Detectable Planet Mass vs. Separation --
using the models of Burrows et al. (2003) and primary star properties from the
literature, 
we can convert our $\Delta$F1 values into a minimum detectable planet mass
for each object.  Objects above the 1400 K methane cutoff line (horizontal
dashed line) are not detected with the SDI device.
For details, see Nielsen et al. 2005 (this conference).}
   \end{figure}

A fully reduced dataset from the VLT SDI device as well as the 
same dataset reduced in a standard AO manner is presented in 
Fig.~\ref{fig:SDIRED}.  This is 40 minutes of data for AB Dor A, a 70 Myr K1V
star at a distance of 14.98 pc (V=6.88).  
Simulated planets have been added at separations of 
0.55, 0.85, and 1.35'' from the primary, with $\Delta$F1 = 10 mag 
(attenuation in magnitudes in the F1 1.575 $\mu$m 
filter) fainter than the primary.
These planets are scaled from unsaturated images of the star 
taken right before the dataset (and have fluxes and photon 
noise in each filter 
appropriate for a T6 object).  In the SDI reduction, the simulated planets are 
detected with S/N $>$ 10 past 0.7''.  In comparison, 
none of the simulated planets are detected in the standard AO data 
reduction and numerous bright super speckles remain in the field.

\section{Contrast Limits and Planet Detectability}

\begin{table}\def~{\hphantom{0}}
  \begin{center}
  \caption{Properties of Example SDI Survey Stars and Comparison Stars}
  \label{tab:properties}
  \begin{tabular}{lcccccccc}\hline
      Case  & Spectral Type  &  Age & Distance & H & V & Exposure Time & $\Delta$F1$^{1}$
& $\Delta$H$^{1}$ \\\hline
	A   &  K2V           &  30 Myr & 45.5 pc & 7.1 & 9.1 & 40 min & 10.5 & 12 \\
	B   &  K1V	     &  70 Myr & 15 pc & 4.8 & 6.9 & 40 min & 10.5 & 12 \\
	C   &  M3V	     &  30 Myr & 24 pc & 7.1 & 12.2 & 40 min & 10 & 11.5 \\ 
10 late K-M stars$^{2}$ & K-M & 0-1 Gyr & 10-50 pc & 6.4-8.7 & 8-12 & 10-25 min & 8.61 & 10.31 \\ 
  Gl 86$^{3}$     &  K1V &  10 Gyr & 10.9 pc & 4.2 & 6.2 & 80 min & 12.8 & 14.3 \\ \hline   
  \end{tabular}
  \end{center}
~$^1$ 5$\sigma$ at 0.5'' ~$^2$ Masciadri et al. IAUC 200 ~$^3$ Mugrauer \& Neuh\"auser 2005
\end{table}

To determine the range of 
star-planet contrasts achievable in our SDI young stars 
survey, we consider three example cases which span the space of our target 
stars: case A -- a high quality dataset (observed with 
seeing of $\sim$0.5''), case B --
AB Dor A, a young solar analogue, and case C, a faint young M star.  Properties
of each example star (distance, age, spectral type, etc.) are presented 
in Table~\ref{tab:properties}.

$\Delta$F1 (5$\sigma$ attenuation in magnitudes in the 1.575 $\mu$m
F1 filter) vs. separation from the primary is presented for each 
of the example cases in Fig.~\ref{fig:deltaH}.
For these datasets, we achieved H band star to planet contrasts $>$25000 
(5$\sigma$ $\Delta$F1(1.575$\mu$m)$>$10 mag, 
$\Delta$H$>$11.5 mag for a T6 spectral type) at a separation of 0.5" from
the primary star -- approaching the photon-noise limit in 40 minutes of data.  

Using the models of Burrows et al. (2003) and adopting values for the primary 
star's age (from the Li 6707 {\AA} line), 
distance, and spectral type from the literature, we 
can convert our measured attenuations for each object into a 
minimum detectable mass (see Nielsen et al. 2005, this conference).  
Minimum detectable mass vs. separation for 
each of the examples is also presented 
in Fig.~\ref{fig:deltaH}.  Although we achieve similar contrast limits
for our example cases (with slightly higher contrasts for brighter targets
as one might expect), the mass and separation of objects detectable
around each varies strongly with age and distance.  
Even though case A was our best quality data,
we are more likely to detect planets for case B and C, simply because these
two objects are closer to the sun, and hence, we 
can resolve the inner $\sim$20 AU around the star.   
For case C, we can detect ($>$5$\sigma$) a 3-5 M$_J$ planet at 
6 AU from the primary.  $\Delta$F1 and $\Delta$H (for a methane object) 
for each survey case as well 
as for other comparison objects from 
independent studies with the VLT SDI device 
are shown in Table~\ref{tab:properties} -- 
it is clear that the achievable contrast varies according
to the magnitude of the object and total exposure time.

To determine what sort of objects we can realistically 
detect with this level of contrast, 
we inserted and then attempted to retrieve simulated T6 dwarf planets to
the case B dataset 
with a variety of separations and $\Delta$F1 contrasts.  $\Delta$F1
contrasts were translated into planet masses using the models of Burrows et 
al. 2003.  In Fig.~\ref{fig:SDIRED}, we plot minimum detectable planet 
mass (for a 5$\sigma$ detection) vs. separation.  For this particular star 
(case B: K1V, 70 Myr, 15 pc), we can detect a 5 M$_{Jup}$ 
planet 12 AU from the star.  In this particular case we were able to detect a 
real 
non-methane companion (AB Dor C) at 3 different epochs and separations from
0.15'' to 0.2'' even though $\Delta$H$>$5 mag (Close et al. this conference
and Nielsen et al. 2005).

\section{Conclusions}\label{sec:concl}

The novel SDI device at the VLT and MMT has been fully commissioned and is 
currently achieving attenuations of $>$25000 ($\Delta$H 
$>$ 11.5 for a T6 spectral type object, $\Delta$F1(1.575$\mu$m)$>$10)
at 0.5''.  With these contrasts, we can detect
a wide range of substellar objects.  For instance, for AB Dor A (a 70 Myr
K1V star 15 pc away) we can detect ($>$5 $\sigma$) a 5 M$_{Jup}$ planet 
12 AU from the star.  For a younger closer star (30 Myr age at 10 pc), we can
detect a 2-4 M$_{Jup}$ planet at 5 AU.

We have currently observed 42 of the youngest ($<$300 Myr), nearest ($<$50 pc)
stars as part of a survey of young, nearby stars. 
We have received time at the VLT for followup observations of 8 
tentative candidates found as part of the survey.  
With a total sample size of 
$\sim$50 stars, we will be able to place strong constraints on the 
frequency and semimajor axis distribution 
of massive extrasolar planets $>$5 AU from their primaries.  From 
scaling laws derived from the distribution of known radial velocity planets
(Marcy et al. 2003, Lineweaver and Grether 2003, Burrows et al. 2003), 
we expect to detect $\sim$4 planets for our total sample
(see Nielsen et al., this conference).
Whether or not we detect planets, 
our survey will begin to measure the true distribution
young massive extrasolar planets $>$5 AU from their primaries and will 
provide valuable contraints for theories of planet formation and migration.   

\begin{acknowledgments}
BAB acknowledges support through the NASA GSRP program.  LMC acknowledges 
support through NSF CAREER and NASA Origins grants.
\end{acknowledgments}

\end{document}